\documentclass[aps,twocolumn,showpacs]{revtex4}
\usepackage{amsmath}
\usepackage{epsfig}

\begin{document}

\title{Further study of the $N\Omega$ dibaryon within constituent quark models}

\author{Hongxia Huang$^1$, Jialun Ping$^1$\footnote{Corresponding
 author: jlping@njnu.edu.cn}, Fan Wang$^2$}

\affiliation{$^1$Department of Physics, Nanjing Normal University,
Nanjing 210097, P.R. China}

\affiliation{$^2$Department of Physics, Nanjing University,
Nanjing 210093, P.R. China}

\begin{abstract}

Inspired by the discovery of the dibaryon $d^{*}$ and the experimental search of
$N\Omega$ dibaryon with the STAR data, we study the strange dibaryon $N\Omega$ further
in the framework of quark delocalization color screening model and chiral quark model.
We have shown $N\Omega$ is a narrow resonance in $\Lambda\Xi$ D-wave scattering before.
However, the $\Lambda$-$\Xi$ scattering data analysis is quite complicated. Here we
calculate the low-energy $N\Omega$ scattering phase shifts, scattering length,
effective range and binding energy to provide another approach of STAR data analysis.
Our results show there exists an $N\Omega$ "bound" state,
which can be observed by the $N$-$\Omega$ correlation analysis with RHIC and LHC data,
or by the new developed automatic scanning system at J-PARC. Besides, we also find
that the hidden color channel-coupling is important for the $N\Omega$ system
to develop intermediate-range attraction.

\end{abstract}


\maketitle

\setcounter{totalnumber}{5}

\section{\label{sec:introduction}Introduction}

The possible existence of dibaryon states was first proposed by F. J. Dyson and
N. Xuong~\cite{Dyson} in 1964. However, this topic got considerable attention
only after R. Jaffe's prediction of the $H$ particle in 1977~\cite{Jaffe}. All
quark models, lattice QCD calculations and other methods, predict that in addition
to $q\bar{q}$ mesons and $q^{3}$ baryons, there should be multiquark systems:
$(q\bar{q})^2$, $q^{4}\bar{q}$, $q^{6}$, quark-gluon hybrids $q\bar{q}g$, $q^{3}g$,
and glueballs~\cite{Jaffe2}. A worldwide theoretical and experimental effort to
search for dibaryon states with and without strangeness has lasted a long time already.

Recently, an $IJ^{P} = 03^{+}$ deep bond $\Delta\Delta$
resonance, called $d^{*}$ in 1989~\cite{dstar}, with a resonance mass $M = 2.37$ GeV and a
width $\Gamma \approx 70$ MeV had been experimentally confirmed by WASA-at COSY
collaboration~\cite{ABC1,ABC2,ABC3}. This ``inevitable" dibaryon has
been extensively studied in our group~\cite{dstar,PRC51,PRC65}. To provide more theoretical input
for experimental search, we included $IJ^{P} = 03^{+}$ $\Delta\Delta$ channel coupling to
the $NN$ $D$-wave scattering phase shifts calculation and a resonance at
$E_{c.m.}=2.36$ MeV had been shown in~\cite{QDCSM2} which called for the precise pn scattering
measurement. Based on the fitting of our model to the $NN$ and $YN$ interaction data
our research had also been extended to dibaryons with strangeness, several strange dibaryon
resonances had been shown in specific baryon-baryon channels~\cite{PRC69,QDCSM3},
among which the most interesting one is the $N\Omega$ state, a very narrow resonance in
$\Lambda\Xi$ D-wave scattering.

In 1987, Goldman {\em et al.} proposed that the $S=-3$, $I=1/2$, $J=2$ dibaryon state
$N\Omega$ might be a narrow resonance in a relativistic quark model \cite{PRL59}.
The quark delocalization color screening model (QDCSM)  confirmed that it was a very
narrow resonance \cite{PRC69}, whereas the chiral quark model also claimed that $N\Omega$
might be bound state \cite{EPJ8}. Recently, the interest
in the $N\Omega$ dibaryon has been revived by lattice QCD calculations of HAL QCD
Collaboration~\cite{HAL}. They reported that the $N\Omega$ is indeed a bound state at
pion mass 875 MeV. To search for $N\Omega$ dibaryon experimentally through the $D-$wave
$\Lambda\Xi$ scattering process is complicated, as neither $\Lambda$ nor $\Xi$ is a stable
particle, $\Xi$ decays to $\Lambda\pi$, and $\Lambda$ decays to $p\pi$. Since $N$ is
stable, it is possible to observe $N\Omega$ scattering experimentally. In order
to provide more choices for experimental search, here we calculate the low-energy scattering
phase shifts, scattering length, effective range and the binding energy of $N\Omega$
system. If it is a sharp resonance, it could be observed by relativistic
heavy-ion collisions data obtained at RHIC and LHC, or by the hadron beam experiments at J-PARC
with their new developed automatic scanning system~\cite{scan}.

Quantum chromodynamics (QCD) has been verified to be the fundamental theory of the strong
interaction in the perturbative region. However, in the low energy region, it is hard to
directly use QCD to study the complicated systems such as hadron-hadron interactions and
multiquark states due to the non-perturbative complication, although lattice QCD has made
impressive progresses on nucleon-nucleon ($NN$) interactions and tetra- and penta-quark
systems~\cite{lqcd,lattice,latt1}. Therefore, various QCD-inspired models have been developed
to get physical insights of the multi-quark systems.

To study the baryon-baryon interaction, the most common approach is the chiral quark model
(ChQM)~\cite{ChQM,Salamanca}, in which the constituent quarks interact with each other through
colorless Goldstone bosons exchange in addition to the colorful one-gluon-exchange and confinement.
To obtain the immediate-range attraction of $NN$ interaction, the chiral partner $\sigma$
meson-exchange has to be introduced. Although the $\sigma$ meson, as $\pi\pi$ $S$-wave resonance,
had been observed by BES collaboration~\cite{BES}, the calculation of the nuclear force with
correlated $\pi\pi$ exchange could not obtain enough attraction~\cite{sigma} as the
phenomenological $\sigma$ meson exchange did.

An alternative approach to study baryon-baryon interaction is the quark delocalization color
screening model (QDCSM), which was developed in 1990s with the aim of explaining the similarities
between nuclear and molecular forces~\cite{QDCSM0}. The model gives a good description of $NN$
and $YN$ interactions and the properties of deuteron~\cite{PRC65,QDCSM1}. It is also employed to
calculate the baryon-baryon scattering phase shifts in the framework of the resonating group
method (RGM), and the dibaryon candidates are also studied with this model~\cite{QDCSM2,QDCSM3}.
Recent studies also show that the $NN$ intermediate-range attraction mechanism in the QDCSM,
quark delocalization and color screening, is equivalent to the $\sigma$ meson-exchange in the
chiral quark model, and the color screening is an effective description of the hidden-color
channels coupling~\cite{QDCSM4,QDCSM5}.

So it is interesting to do a comparative study of the $N\Omega$ system with these two quark models.
This state might serve as a test of the flavor-dependent $q-q$ interaction due to Goldstone-boson
exchange and due to quark delocalization  color screening, because there is no common flavor quark
between $N$ and $\Omega$ and so no quark exchange between these two baryons.

The structure of this paper is as follows. A brief introduction of
constituent quark models used is given in section II. Section III devotes to the numerical
results and discussions. The summary is shown in the last section.

\section{Two quark models}

In order to estimate the model dependence of dibaryon prediction, the two quark models:
ChQM and QDCSM, are employed here to study the $N\Omega$ system.

\subsection{Chiral quark  model}

The Salamanca model was chosen as the representative of the chiral quark models, because
the Salamanca group's work covers the hadron spectra, nucleon-nucleon interaction, and
multiquark states. In this model,the constituent quarks interact with each other through
Goldstone boson exchange and one-gluon-exchange in addition to the color confinement.
The model details can be found in Ref.~\cite{Salamanca}. Here we only give the Hamiltonian:
\begin{widetext}
\begin{eqnarray}
H & = & \sum_{i=1}^6\left(m_i+\frac{p_i^2}{2m_i}\right)-T_{CM} +\sum_{j>i=1}^6
\left(V^{C}_{ij}+V^{G}_{ij}+V^{\chi}_{ij}+V^{\sigma}_{ij}\right), \\
V^{C}_{ij} & = & -a_{c} \boldsymbol{\lambda}^c_{i}\cdot \boldsymbol{
\lambda}^c_{j} ({r^2_{ij}}+v_{0}), \label{sala-vc} \\
V^{G}_{ij} & = & \frac{1}{4}\alpha_s \boldsymbol{\lambda}^{c}_i \cdot
\boldsymbol{\lambda}^{c}_j
\left[\frac{1}{r_{ij}}-\frac{\pi}{2}\delta(\boldsymbol{r}_{ij})(\frac{1}{m^2_i}+\frac{1}{m^2_j}
+\frac{4\boldsymbol{\sigma}_i\cdot\boldsymbol{\sigma}_j}{3m_im_j})-\frac{3}{4m_im_jr^3_{ij}}
S_{ij}\right] \label{sala-vG} \\
V^{\chi}_{ij} & = & V_{\pi}( \boldsymbol{r}_{ij})\sum_{a=1}^3\lambda
_{i}^{a}\cdot \lambda
_{j}^{a}+V_{K}(\boldsymbol{r}_{ij})\sum_{a=4}^7\lambda
_{i}^{a}\cdot \lambda _{j}^{a}
+V_{\eta}(\boldsymbol{r}_{ij})\left[\left(\lambda _{i}^{8}\cdot
\lambda _{j}^{8}\right)\cos\theta_P-(\lambda _{i}^{0}\cdot
\lambda_{j}^{0}) \sin\theta_P\right] \label{sala-Vchi1} \\
V_{\chi}(\boldsymbol{r}_{ij}) & = & {\frac{g_{ch}^{2}}{{4\pi
}}}{\frac{m_{\chi}^{2}}{{\
12m_{i}m_{j}}}}{\frac{\Lambda _{\chi}^{2}}{{\Lambda _{\chi}^{2}-m_{\chi}^{2}}}}%
m_{\chi} \left\{(\boldsymbol{\sigma}_{i}\cdot
\boldsymbol{\sigma}_{j})
\left[ Y(m_{\chi}\,r_{ij})-{\frac{\Lambda_{\chi}^{3}}{m_{\chi}^{3}}}%
Y(\Lambda _{\chi}\,r_{ij})\right] \right.\nonumber \\
&& \left. +\left[H(m_{\chi}
r_{ij})-\frac{\Lambda_{\chi}^3}{m_{\chi}^3}
H(\Lambda_{\chi} r_{ij})\right] S_{ij} \right\}, ~~~~~~\chi=\pi, K, \eta, \\
V^{\sigma}_{ij} & = & -{\frac{g_{ch}^{2}}{{4\pi }}}
{\frac{\Lambda _{\sigma}^{2}}{{\Lambda _{\sigma}^{2}-m_{\sigma}^{2}}}}%
m_{\sigma}\left[ Y(m_{\sigma}\,r_{ij})-{\frac{\Lambda _{\sigma}}{m_{\sigma}}}%
Y(\Lambda _{\sigma}\,r_{ij})\right] , \\
S_{ij}&=&\left\{ 3\frac{(\boldsymbol{\sigma}_i
\cdot\boldsymbol{r}_{ij}) (\boldsymbol{\sigma}_j\cdot
\boldsymbol{r}_{ij})}{r_{ij}^2}-\boldsymbol{\sigma}_i \cdot
\boldsymbol{\sigma}_j\right\},\\
H(x)&=&(1+3/x+3/x^{2})Y(x),~~~~~~
 Y(x) =e^{-x}/x. \label{sala-vchi2}
\end{eqnarray}
\end{widetext}
Where $\alpha_s$ is the quark-gluon coupling constant. In order to
cover the wide energy scale from light, strange to heavy quark one
introduces an effective scale-dependent quark-gluon coupling
constant $\alpha_s(\mu)$~\cite{JPG31},
\begin{equation}
\alpha_s(\mu)=\frac{\alpha_{0}}{\ln\left(\frac{\mu^{2}+\mu_0^{2}}{\Lambda_0^{2}}\right)},
\label{alpha-s}
\end{equation}
where $\mu$ is the reduced mass of the interacting quark-pair. The coupling constant
$g_{ch}$ for chiral field is determined from the $NN\pi$ coupling constant through
\begin{equation}
\frac{g_{ch}^{2}}{4\pi }=\left( \frac{3}{5}\right) ^{2}{\frac{g_{\pi NN}^{2}%
}{{4\pi }}}{\frac{m_{u,d}^{2}}{m_{N}^{2}}}\label{gch}.
\end{equation}
The other symbols in the above expressions have their usual meanings.

For dibaryon with strangeness, Two versions of chiral quark model~\cite{Garcilazo,QBLi}
had been used. One is the so called extended chiral SU(2) quark model,
in which $\sigma$ meson-exchange was used between any quark pair.
Another is the chiral SU(3) quark model, where full SU(3) scalar octet meson-exchange was used.
In order to analyze the effect of the scalar meson-exchange of the ChQM, in this work,
we use three kinds of chiral quark models: (1) SU(2) ChQM1, where $\sigma$ meson-exchange is universal,
i.e., it exchanges between any quark pair; (2) SU(2) ChQM2, where $\sigma$ meson is restricted to
exchange between $u$ and/or $d$ quark pair only; (3) SU(3) ChQM, where the full SU(3) scalar octet
meson-exchange is employed. These scalar potentials have the same functional form as the one of
SU(2) ChQM but a different SU(3) operator dependence~\cite{Garcilazo}, that is,
\begin{eqnarray}
V^{\sigma_{a}}_{ij} & = & V_{a_{0}}(
\boldsymbol{r}_{ij})\sum_{a=1}^3\lambda _{i}^{a}\cdot \lambda
_{j}^{a}+V_{\kappa}(\boldsymbol{r}_{ij})\sum_{a=4}^7\lambda
_{i}^{a}\cdot \lambda _{j}^{a} \nonumber \\
& & +V_{f_{0}}(\boldsymbol{r}_{ij})\lambda _{i}^{8}\cdot \lambda
_{j}^{8}+V_{\sigma}(\boldsymbol{r}_{ij})\lambda _{i}^{0}\cdot
\lambda _{j}^{0} \label{sala-su3} \\
V_{k}(\boldsymbol{r}_{ij}) & = & -{\frac{g_{ch}^{2}}{{4\pi }}}
{\frac{\Lambda _{k}^{2}m_{k}}{{\Lambda_{k}^{2}-m_{k}^{2}}}}%
\left[ Y(m_{k}\,r_{ij})-{\frac{\Lambda _{k}}{m_{k}}}%
Y(\Lambda _{k}\,r_{ij})\right] , \nonumber
\end{eqnarray}
with $k=a_{0}, \kappa, f_{0}$ or $\sigma$.

\subsection{Quark delocalization color screening model}

The Hamiltonian of QDCSM is almost the same as that of chiral quark model
but with two modifications~\cite{QDCSM0,QDCSM1}: First, there
is no $\sigma$-meson exchange in QDCSM, and second, the screened
color confinement is used between quark pairs reside in different
baryon orbits. That is
\begin{equation}n
V_{ij}^{C}=\left \{
\begin{array}{ll}
-a_{c}\boldsymbol{\mathbf{\lambda}}^c_{i}\cdot
\boldsymbol{\mathbf{ \lambda}}^c_{j}~(r_{ij}^2+ v_0) &
  \mbox{if \textit{i},\textit{j} in the same} \\
  &  \mbox{baryon orbit} \\
-a_{c}\boldsymbol{\mathbf{\lambda}}^c_{i}\cdot
\boldsymbol{\mathbf{
\lambda}}^c_{j}~(\frac{1-e^{-\mu_{ij}\mathbf{r}_{ij}^2}}{\mu_{ij}}+
v_0) & \mbox{otherwise}
\end{array}
\right.\label{QDCSM-vc}
\end{equation}
The color screening constant $\mu_{ij}$ in Eq.(\ref{QDCSM-vc}) is
determined by fitting the deuteron properties, $NN$ scattering
phase shifts and $N\Lambda$, $N\Sigma$ scattering cross sections,
$\mu_{uu}=0.45$, $\mu_{us}=0.19$ and
$\mu_{ss}=0.08$, which satisfy the relation, $\mu_{us}^{2}=\mu_{uu}*\mu_{ss}$.

The quark delocalization in QDCSM is realized by replacing the
left, right centered single Gaussian functions, the single
particle orbital wave function in usual quark cluster model
\begin{eqnarray}
\phi_\alpha(\boldsymbol {S_{i}})=\left(\frac{1}{\pi
b^2}\right)^{\frac{3}{4}}e^ {-\frac{(\boldsymbol {r}-\boldsymbol
{S_i}/2)^2}{2b^2}},
 \nonumber\\
\phi_\beta(-\boldsymbol {S_{i}})=\left(\frac{1}{\pi
b^2}\right)^{\frac{3}{4}}e^ {-\frac{(\boldsymbol {r}+\boldsymbol
{S_i}/2)^2}{2b^2}} .
 \
\end{eqnarray}
with delocalized ones,

\begin{eqnarray}
{\psi}_{\alpha}(\boldsymbol {S_{i}},\epsilon) &=&
\left({\phi}_{\alpha}(\boldsymbol{S_{i}})
+\epsilon{\phi}_{\alpha}(-\boldsymbol{S_{i}})\right)/N(\epsilon),
\nonumber \\
{\psi}_{\beta}(-\boldsymbol {S_{i}},\epsilon) &=&
\left({\phi}_{\beta}(-\boldsymbol{S_{i}})
+\epsilon{\phi}_{\beta}(\boldsymbol{S_{i}})\right)/N(\epsilon),
\nonumber \\
N(\epsilon)&=&\sqrt{1+\epsilon^2+2\epsilon e^{{-S}_i^2/4b^2}}.
\end{eqnarray}
The delocalization parameter $\epsilon(\boldsymbol{S}_i)$ is
determined by the dynamics of the quark system rather than
adjusted parameters. In this way, the system can choose its most
favorable configuration through its own dynamics in a larger
Hilbert space.

The parameters of these models are given in
Table~\ref{parameters}. The calculated baryon masses in comparison with experimental values are shown in Table~\ref{mass}.

\begin{table}[ht]
\caption{\label{parameters}The parameters of two models:
$m_{\pi}=0.7$fm$^{-1}$, $m_{ k}=2.51$fm$^{-1}$,
$m_{\eta}=2.77$fm$^{-1}$, $m_{\sigma}=3.42$fm$^{-1}$,
$m_{a_{0}}=m_{\kappa}=m_{f_{0}}=4.97$fm$^{-1}$,
$\Lambda_{\pi}=4.2$fm$^{-1}$, $\Lambda_{K}=5.2$fm$^{-1}$,
$\Lambda_{\eta}=5.2$fm$^{-1}$, $\Lambda_{\sigma}=4.2$fm$^{-1}$,
$\Lambda_{a_{0}}=\Lambda_{\kappa}=\Lambda_{f_{0}}=5.2$fm$^{-1}$,
$g_{ch}^2/(4\pi)$=0.54, $\theta_p$=$-15^{0}$. }
\begin{tabular}{llllll}
\hline\hline
              &                         &QDCSM     &SU(2)ChQM1  &SU(2)ChQM2  &SU(3)ChQM\\   \hline
              &$b$(fm)                  &~~0.518     &~~0.518       &~~0.518       &~~0.518    \\
              & $m_u$(MeV)              &~~313       &~~313         &~~313         &~~313      \\
              & $m_d$(MeV)              &~~313       &~~313         &~~313         &~~313      \\
              & $m_s$(MeV)              &~~573       &~~573         &~~536         &~~573      \\
                                                                                           \hline
              &$a_c$(MeV)               &~~58.03     &~~48.59       &~~48.59       &~~48.59    \\
              &$\mu_{uu}$(fm$^{-2}$)    &~~0.45      &~~-           &~~-           &~~-        \\
              &$\mu_{us}$(fm$^{-2}$)    &~~0.19      &~~-           &~~-           &~~-        \\
              &$\mu_{ss}$(fm$^{-2}$)    &~~0.08      &~~-           &~~-           &~~-        \\
              &$v_{0}$(MeV)             &-1.2883   &-1.2145       &-1.2145       &-0.961\\
                                                                       \hline
              &$\alpha_0$               &~~0.510     &~~0.510       &~~0.510      &~~0.583     \\
              &$\Lambda_0($fm$^{-1})$   &~~1.525     &~~1.525       &~~1.525      &~~1.616     \\
              &$\mu_0$(MeV)             &~~445.808   &445.808       &445.808      &422.430     \\
                                                                       \hline

\hline\hline
\end{tabular}
\end{table}

\begin{table}[ht]
\caption{\label{mass}The masses of ground-state baryons(in MeV).}
\begin{tabular}{lcccccccc}
\hline \hline
               & ~~$N$~~            & ~~$\Delta$~~ & ~~$\Lambda$~~ & ~~$\Sigma$~~
               & ~~$\Sigma^*$~~     & ~~$\Xi$~~    & ~~$\Xi^*$~~   & ~~$\Omega$~~  \\ \hline
QDCSM          & 939          & 1232       & 1124    & 1238
                & 1360         & 1374       & 1496   & 1642       \\ \hline
SU(2)ChQM1      & 939          & 1232       & 1124     & 1239
                & 1360         & 1376       & 1498     & 1644     \\\hline
SU(2)ChQM2      & 939          & 1232       & 1137     & 1245
                & 1376         & 1375       & 1506     & 1620     \\\hline
SU(3)ChQM      & 939          & 1232       & 1123     & 1267
                & 1344         & 1398       & 1475     & 1625     \\\hline
 Expt  & 939  &1232 &1116 &1193 &1385 &1318 &1533 &1672\\  \hline\hline
\end{tabular}
\end{table}

\section{The results and discussions}

Here, we investigate the low-energy properties of the $N$-$\Omega$ scattering with
quantum numbers $S,I,J=-3,1/2,2$ within various constituent quark models mentioned above.
The effects of channel-coupling are studied carefully, both color-singlet channels
(the color symmetry of $3q$-cluster is $[111]$) and hidden-color channels (the color
symmetry of $3q$-cluster is $[21]$, which is called colorful $3q$-cluster) are included.
The colorful $3q$-cluster are listed in Table~\ref{colorful} with color symmetry $[c]$,
spin symmetry $[\sigma]$, flavor symmetry $[f]$, isospin $I$, and strangeness $S$.
The orbital symmetry is limited to be $[3]$. The labels of all 16 coupled channels of
the $N\Omega$ system are listed in Table \ref{channels}.

To check whether or not there is a bound $N\Omega$ state, a dynamic calculation
based on the resonating group method (RGM)~\cite{RGM} has been done. Expanding
the relative motion wavefunction between two clusters in the RGM equation by Gaussian bases,
the integro-differential equation of RGM reduces to algebraic equation-a generalized
eigen-equation. The energy of the system is obtained by solving this generalized
eigen-equation. In the calculation, the baryon-baryon separation is taken to be less
than 6 fm (to keep the dimensions of matrix manageably small). The binding energies of
$N\Omega$ state in various quark models are listed in Table~\ref{bound}, where $B_{sc}$
stands for the binding energy of the single channel $N\Omega$, $B_{5cc}$ means
the binding energy with the five color-singlet channels included, and $B_{16cc}$ refers
to the binding energy with all the 16 channels coupling.
\begin{widetext}
\begin{center}
\begin{table}[ht]
\caption{\label{colorful}The symmetries of colorful $3q-$cluster.}
\begin{tabular}{lcccccccccccc}
\hline \hline
        & ~~$N^{\prime}$~~           & ~~$\Sigma^{\prime}$~~ & ~~$\Xi^{\prime}$~~ &
        ~~$\Lambda^{\prime}$~~       & ~~$\Sigma^{*\prime}$~~     & ~~$\Xi^{*\prime}$~~  &
        ~~$\Omega^{\prime}$~~  & ~~$N^{"}$~~   & ~~$\Sigma^{"}$~~ & ~~$\Xi^{"}$~~ &
        ~~$\Lambda^{"}$~~ & ~~$\Lambda_{s}^{\prime}$~~ \\ \hline
$[c]$   & $[21]$      & $[21]$       & $[21]$     & $[21]$  & $[21]$    &  $[21]$
     & $[21]$     & $[21]$ & $[21]$    & $[21]$     & $[21]$     & $[21]$     \\ \hline
$[\sigma]$     & $[21]$       & $[21]$   & $[21]$     & $[21]$  & $[21]$
        & $[21]$   & $[21]$  & $[3]$    & $[3]$      & $[3]$       & $[3]$     & $[21]$     \\\hline
$[f]$      & $[21]$    & $[21]$       & $[21]$     & $[21]$& $[3]$
  & $[3]$  & $[3]$     & $[21]$& $[21]$          & $[21]$       & $[21]$     & $[111]$    \\\hline
$I$      & $\frac{1}{2}$        & 1    & $\frac{1}{2}$     & 0
         & 1        & $\frac{1}{2}$    & 0   & $\frac{1}{2}$  &  1  & $\frac{1}{2}$  &  0  &  0     \\\hline
$S$  & 0  & 1 &  2 & 1 &1 &2 &3 &0 & 1 &2 &1 &1 \\  \hline\hline
\end{tabular}
\end{table}
\end{center}
\end{widetext}

\begin{table}[ht]
\caption{Channels of the $N\Omega$ system.}
\begin{tabular}{cccccccc}
\hline \hline
 ~~ 1 & ~~2 & ~~3 & ~~4 & ~~5 & ~~6 & ~~7 & ~~8  \\ \hline
 ~~$\Xi^{*}\Sigma$ & ~~$\Xi\Sigma^{*}$
                  & ~~$\Xi^{*}\Lambda$ & ~~$N\Omega$
                  & ~~$\Xi^{*}\Sigma^{*}$  & ~~$\Xi^{"}\Sigma^{*\prime}$
                  & ~~$\Xi^{*\prime}\Sigma^{"}$ & ~~$\Xi^{"}\Sigma^{"}$  \\  \hline
   ~~9 & ~~10 & ~~11 & ~~12 & ~~13 & ~~14 & ~~15 &~~16  \\ \hline
 ~~$\Xi^{"}\Sigma$ & ~~$N^{\prime}\Omega^{\prime}$
                  & ~~$\Xi^{*\prime}\Lambda^{"}$ & ~~$\Xi^{\prime}\Lambda^{"}$
                  & ~~$\Xi^{"}\Lambda^{"}$  & ~~$\Xi^{"}\Lambda_{s}^{\prime}$
                  & ~~$\Xi^{\prime}\Sigma^{"}$ & ~~$\Xi^{"}\Lambda^{\prime}$  \\  \hline
  \hline
\end{tabular}
\label{channels}
\end{table}

\begin{table}[ht]
\caption{The binding energies B with channel-coupling.}
\begin{tabular}{lcccc}
\hline \hline
  & ~$B_{sc}~(MeV)$~ & ~$B_{5cc}~(MeV)$~ & ~$B_{16cc}~(MeV)$~    \\ \hline
QDCSM & ub  & -6.4 & --   \\ \hline
 SU(2)ChQM1 & -19.6 & -48.8 & -119.5   \\ \hline
 SU(2)ChQM2 & ub & ub & -38.3   \\ \hline
 SU(3)ChQM & ub & ub & -13.7   \\ \hline
  \hline
\end{tabular}
\label{bound}
\end{table}

The single channel calculation shows that the $N\Omega$ is unbound (labelled as 'ub'
in Table~\ref{bound}) in all quark models except the SU(2)ChQM1, in which a universal
$\sigma$ meson-exchange is used. To show the contribution of each interaction term
to the energy of the system, effective potentials between $N$ and $\Omega$ 
are shown in Fig.~1, in which the contributions from various terms, the kinetic energy
($V_{vk}$), the confinement ($V_{con}$), the one-gluon-exchange ($V_{oge}$), the
one-boson-exchange ($V_{\pi}$, $V_{K}$, and $V_{\eta}$), and the scalar octet
meson-exchange ($V_{\sigma}$, $V_{a_{0}}$, $V_{\kappa}$, and $V_{f_{0}}$) are given.
For the ChQM, the confinement, the one-gluon-exchange do not contribute to the effective
potential between $N$ and $\Omega$, because no quark exchange between these two baryons,
the pion and the $a_(0)$ meson do not contribute either because they do not exchange
between u(d) and s quark. The contributions of other terms to the effective potential
are shown in Fig. 1(a), Fig. 1(b) and Fig. 1(c), in which we can see that the attraction
of the $N\Omega$ system mainly comes from the $\sigma$ meson-exchange interaction.
In SU(2)ChQM1, the attraction from $\sigma$ meson-exchange is so large that it leads to
a deep attractive potential between the N and $\Omega$, which makes the $N\Omega$ bound.
While in SU(2)ChQM2, where the $\sigma$ meson is restricted to exchange between the $u$ and $d$
quarks only, there is no $\sigma$ meson-exchange interaction between $N$ and $\Omega$.
So the total potential is repulsive, resulting in unbound $N\Omega$.
In SU(3)ChQM, even though the universal $\sigma$ meson-exchange introduces large attraction,
which is canceled by the repulsive potentials of $\kappa$ and $f_{0}$ meson-exchange,
and also causes the $N\Omega$ unbound.

\begin{figure*}
\epsfxsize=4.8in \epsfbox{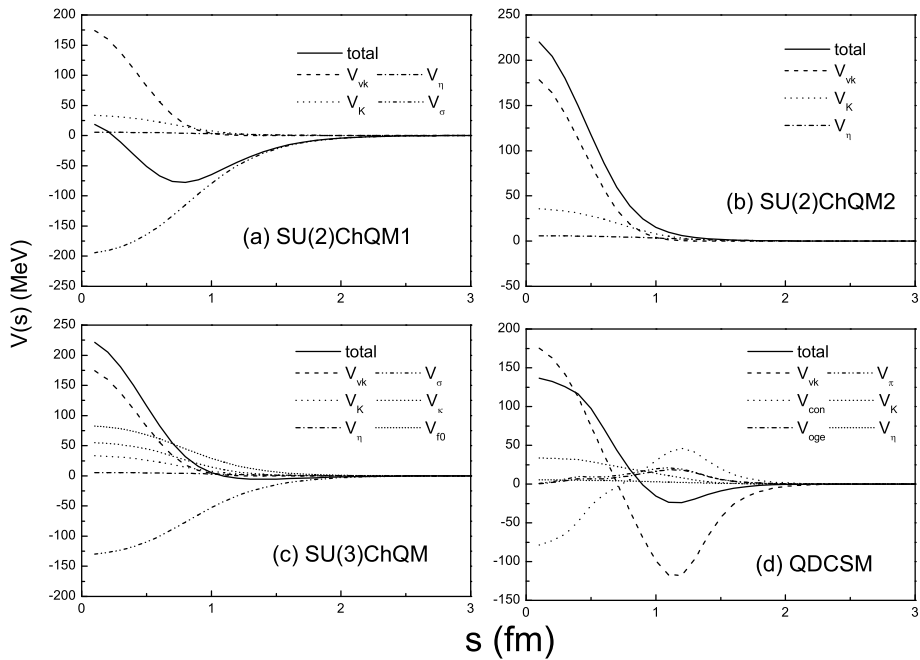} \vspace{-0.1in}

\caption{The contributions to the effective potential from various
terms of interactions.}
\end{figure*}

Things are different in QDCSM. In this model, quark delocalization and color
screening work together to provide short-range repulsion and intermediate-range
attraction. We illustrate this mechanism by showing contributions of all interaction terms
to the effective potential in Fig. 1(d), from which we see that the attraction of the $N\Omega$
system mainly comes from the kinetic energy term due to quark delocalization, other terms provide
repulsive potentials, which reduce the total attraction of the $N\Omega$ potential. The single
channel approximation of the $N\Omega$ in QDCSM is also unbound as shown in Table~\ref{bound}.

Then we consider the effects of channel-coupling. From Table~\ref{bound} we can see that in
SU(2)ChQM1, the binding energy of the $N\Omega$ increases to 48.8 MeV by including
color-singlet channels coupling, and reaching the largest binding energy of $-119.5$ MeV
with the hidden-color channels coupling, which shows the channel-coupling has a huge influence
on the $N\Omega$ state. In the SU(2)ChQM2 and SU(3)ChQM, the effect of color-singlet channels
coupling is not large enough to make the $N\Omega$ bound, only additional hidden-color channels
coupling leads to the bound $N\Omega$ state in both models. In QDCSM, since it contains
hidden-color channels coupling effect already through the color screening~\cite{QDCSM2,QDCSM5},
including the color-singlet channels coupling only already makes the $N\Omega$ state bound.
All these results are consistent with our previous study of non-strange channels~\cite{QDCSM2,QDCSM5},
in which we found the QDCSM with color-singlet channels copling only had similar results of
the chiral quark models with both color-singlet and hidden-color channels coupling.

If the $N\Omega$ dibaryon is an $S-$wave bound state, the strong decays to $S-$wave
octet-decuplet ($\Sigma\Xi^{*}$, $\Xi\Sigma^{*}$, and $\Lambda\Xi^{*}$) and decuplet-decuplet
($\Sigma^{*}\Xi^{*}$) channels are prohibited kinematically because of its mass will be lower than the
thresholds of these channels. It can only decay to the $D-$wave $\Lambda\Xi$. We had done a scattering
calculation by coupling the $S-$wave $N\Omega$ and the $D-$wave $\Lambda\Xi$, and found
the $S-$wave bound state $N\Omega$ showed as a resonance in the $D-$wave $\Lambda\Xi$ scattering
process~\cite{QDCSM3}. Because of only the tensor interaction can couple the $S-$wave bound state
to $D-$wave octet-octet baryon channel, so the energy of the bound state are pushed down only a little
and the resonance width is generally small. We learnt from the Shanghai group of STAR collaboration
that it is quite complicate to analyze the $D-$wave $\Lambda\Xi$ scattering data because both $\Lambda$
and $\Xi$ are weak interaction unstable.To provide more theoretical input, here we calculate
the low-energy scattering phase shifts, scattering length and the effective range of
$N\Omega$ system. All results given below are calculated with 16 channels coupling in the chiral
quark models and 5 color-singlet channels coupling in QDCSM.

Firstly, we calculate the $S-$wave $N\Omega$ low-energy scattering phase shifts by using the well
developed Kohn-Hulthen-Kato(KHK) variational method. The details can be found in Ref.~\cite{RGM}.
Fig. 2 illustrates the scattering phase shifts of the $S-$wave $N\Omega$. It is obvious that in all
four quark models, the scattering phase shifts go to 180 degrees at $E_{c.m.}\sim 0$ and rapidly
decreases as $E_{c.m.}$ increases, which implies the existence of a bound state. The results are
consistent with the the bound state calculation shown before andthe lattice QCD calculation~\cite{HAL}.

\begin{figure}
\epsfxsize=3.0in \epsfbox{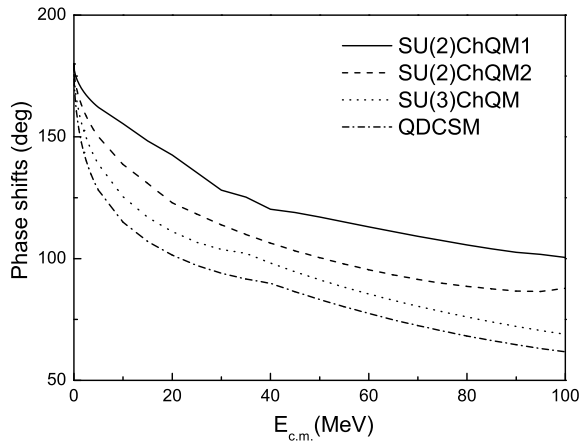} \vspace{-0.1in}

\caption{The phase shifts of $S-$wave $N\Omega$ dibaryon.}
\end{figure}

Secondly, by using these low-energy phase shifts, we extract the scattering length $a_{0}$ and
the effective range $r_{0}$ of $N\Omega$ scattering. Then the binding energy $B^{\prime}$
is obtained. The scattering length and effective range are calculated from the low-energy
scattering phase shifts as:
\begin{eqnarray}
kcot\delta & = & -\frac{1}{a_{0}}+\frac{1}{2}r_{0}k^{2}+{\cal
O}(k^{4})
\end{eqnarray}
The binding energy $B^{\prime}$ is calculated according to the relation:
\begin{eqnarray}
B^{\prime} & = &\frac{\hbar^2\alpha^2}{2\mu}
\end{eqnarray}
where $\mu$ is the reduced mass of the $N\Omega$; $\alpha$ is the wave number which can be obtained
from the relation~\cite{Babenko}:
\begin{eqnarray}
r_{0} & = &\frac{2}{\alpha}(1-\frac{1}{\alpha a_{0}})
\end{eqnarray}
Please note that we use another method to calculate the binding energy, labelled as $B^{\prime}$.
The results are listed in Table \ref{length}.
\begin{center}
\begin{table}[h]
\caption{The scattering length $a_{0}$, effective range $r_{0}$,
and binding energy $B^{'}$ of the $N\Omega$ dibaryon.}
\begin{tabular}{lcccc}
\hline \hline
  & ~$a_{0}~(fm)$~ & ~$r_{0}~(fm)$~ & ~$B^{'}~(MeV)$~    \\ \hline
 QDCSM & 2.8007  & 0.5770 & -5.2    \\ \hline
 SU(2)ChQM1 & 0.8103 & 0.3609 & -110.3   \\ \hline
 SU(2)ChQM2 & 1.3808  & 0.6018 & -37.3    \\ \hline
 SU(3)ChQM & 1.9870 & 0.7064 & -13.7   \\ \hline
  \hline
\end{tabular}
\label{length}
\end{table}
\end{center}
From Table~\ref{length}, we can see that in four quark models, the scattering length are all positive,
which implies the existence of a bound state of $N\Omega$. The binding energies of the
$N\Omega$ dibaryon from the two methods (see $B$ and $B^{'}$) are coincident with each other.

\section{Summary}
In quark model, the hadron interaction usually depends critically
upon the contribution of the color-magnetic interaction. In this letter, we show
that the $N$-$\Omega$ system with quantum numbers $SIJ=-3\frac122$ their effective interaction have
very small contribution from the color-magnetic interaction due to its special quark content.
Because of this, the study of this state might teach us something about the mechanism of the
intermediate-range attraction of the baryon-baryon interaction. In $NN$ case we had shown that the
phenomenological $\sigma$ meson exchange is equivalent to the quark delocalization
and color screening and the latter is an effective description of hidden color channel coupling.
For the $N\Omega$ system, the chiral quark model now only has the Goldstone boson exchange,
which will predict there is not a bound state if the
$\sigma$-meson is not universal exchange between any quark pair. The physical observed $\sigma$
meson is an $u\bar{u}+d\bar{d}$ system which should not exchange between $u(d)$ and $s$ quark.
Only if one includes the hidden color channels coupling then the chiral quark model can
accomodate a bound $N\Omega$ state. On the other hand, the QDCSM predicts there will be a bound
$N\Omega$ state if the color singlet channels coupling has been taken into account. The quark
delocalization and color screening with the five color singlet channels coupling provides enough
effective attraction to bound the $N\Omega$.
Therefore if experiment confirms the existence of $N\Omega$ dibaryon state, bound or appears as a
D-wave resonance in $\Lambda$-$\Xi$ scattering, It will be a signal showing that the quark delocalization
and color screening (an effective description of hidden color channels coupling) is really responsible
for the intermediate range attraction of baryon-baryon interaction. This mechanism is also preferred
by the similarity between nuclear force and molecular force.

Experimental confirmation of the $N\Omega$ dibaryon will provide the second sample of six-quark system.
It will provide another sample for the persue of low energy scale QCD property and we wish there will be
more experimental groups to be involved in the search of $N\Omega$ dibaryon with quantum numbers
$SIJ^P=-3\frac122^+$.

\acknowledgments{This work is supported partly by the National Science Foundation of China under
Contract Nos. 11175088, 11035006, 11205091}


\begin{thebibliography}{99}
\bibitem{Dyson} F. J. Dyson and N. H. Xuong, Phys. Rev. Lett. {\bf 13}, 815 (1964).
\bibitem{Jaffe} R. L. Jaffe, Phys. Rev. Lett. {\bf 38}, 195 (1977).
\bibitem{Jaffe2} R. L. Jaffe, Phys. Rep. {\bf 409}, 1 (2005); F.
E. Close, Int. J. Mod. Phys. A {\bf 20}, 5156 (2005).
\bibitem{dstar} T. Goldman, K. Maltman, G. J. Stephenson, K. E. Schmidt and F. Wang, Phys. Rev. C
{\bf 39}, 1889 (1989).
\bibitem{ABC1} M. Bashkanov {\em et al} (CELSIUS-WASA Collaboration),
Phys. Rev. Lett. {\bf 102}, 052301 (2009).
\bibitem{ABC2} P. Adlarson {\em et al} (WASA-at-COSY Collaboration),
Phys. Rev. Lett. {\bf 106}, 242302 (2011).
\bibitem{ABC3}P. Adlarson et al. (WASA-at-COSY collaboration), Phys. Rev. Lett. {\bf 112}, 202301 (2014).
\bibitem{PRC51} F. Wang, J. L. Ping, G. H. Wu, L. J. Teng and T. Goldman,
 Phys. Rev. C {\bf 51}, 3411 (1995).
\bibitem{PRC65} J. L. Ping, H. R. Pang, F. Wang and T. Goldman, Phys. Rev. C {\bf 65}, 044003 (2002).
\bibitem{QDCSM2} J. L. Ping, H. X. Huang, H. R. Pang, F. Wang and C. W. Wong, Phys. Rev.
C {\bf 79}, 024001 (2009).
\bibitem{PRC69} H. R. Pang, J. L. Ping, F. Wang, T. Goldman, and E. G. Zhao,
 Phys. Rev. C {\bf 69}, 065207 (2004).
\bibitem{QDCSM3} M. Chen, H. X. Huang, J. L. Ping and F. Wang, Phys. Rev. C {\bf 83}, 015202 (2011).
\bibitem{PRL59} T. Goldman, K. Maltman, G. J. Stephenson, K. E. Schmidt and F. Wang,
Phys. Rev. Lett. {\bf 59}, 627 (1987).
\bibitem{EPJ8} Q. B. Li, P. N. Shen, Eur. Phys. J. A {\bf 8}, 417 (2000).
\bibitem{HAL} F. Etminan, H. Nemura, S. Aoki, {\em et al.}, arXiv:1403.7284v2.
\bibitem{scan} J. Yoshida et al., talk in International Workshop on Hadron Nuclear Physics 2015,
July 7-11, Krabi, Thailand, p.39 of Book of Abstracts HNP2015.
\bibitem{lqcd} N. Ishii, S. Aoki and T. Hatsuda, Phys. Rev. Lett. {\bf 99} 022001 (2007).
\bibitem{lattice} C. Alexandrou, P. de Forcrand, and A. Tsapalis, Phys. Rev. D
 {\bf 65}, 054503 (2002); T. T. Takahashi, H. Suganuma, Y. Nemoto, and H. Matsufuru, Phys.
Rev. D {\bf 65}, 114509 (2002).
\bibitem{latt1} F. Okiharu, H. Suganuma, T. T. Takahashi, Phys. Rev. Lett. \textbf{94},
192001 (2005).
\bibitem{ChQM} Y. Fujiwara, C. Nakamoto and Y. Suzuki, Phys. Rev. Lett. {\bf 76}, 2242 (1996);
Y. W. Yu, Z. Y. Zhang, P. N. Shen and L. R. Dai, Phys. Rev. C {\bf 52}, 3393 (1995).
\bibitem{Salamanca} A. Valcarce, H. Garcilazo, F. Fern\'{a}ndez and P. Gonzalez,
  Rep. Prog. Phys. {\bf 68}, 965 (2005) and references therein.
\bibitem{BES} M. Ablikim, et al. [BES Collaboration], Phys. Lett. B {\bf 598}, 149 (2004).
\bibitem{sigma} N. Kaiser, S. Grestendorfer and W. Weise, Nucl.
Phys. {\bf A637}, 395 (1998); E. Oset, H. Toki, M. Mizobe and T. T.
Takahashi, Prog. Theo. Phys. {\bf 103}, 351 (2000); M. M. Kaskulov
and H. Clement, Phys. Rev. {\bf C70}, 014002 (2004).
\bibitem{QDCSM0} F. Wang, G. H. Wu, L. J. Teng and T. Goldman, Phys. Rev. Lett. {\bf 69}, 2901 (1992);
G. H. Wu, L. J. Teng, J. L. Ping, F. Wang and T. Goldman, Phys. Rev. C {\bf
53}, 1161 (1996).
\bibitem{QDCSM1} J. L. Ping, F. Wang and T. Goldman, Nucl. Phys. A {\bf 657},
95 (1999); G. H. Wu, J. L. Ping, L. J. Teng {\em et al.}, Nucl. Phys. A {\bf 673},
279 (2000); H. R. Pang, J. L. Ping, F. Wang and T. Goldman, Phys. Rev. C {\bf 65},
014003 (2001).
\bibitem{QDCSM4} L. Z. Chen, H. R. Pang, H. X. Huang, J.
L. Ping and F. Wang, Phys. Rev. C {\bf 76}, 014001 (2007);
\bibitem{QDCSM5} H. X. Huang, P. Xu, J. L. Ping and F. Wang, Phys. Rev. C {\bf 84}, 064001 (2011).
\bibitem{JPG31} J. Vijande, F. Fernandez and A. Valcarce, J. Phys. G {\bf 31}, 481 (2005).
\bibitem{Garcilazo} H. Garcilazo, T. F. Carames and A. Valcarce, Phys. Rev. C {\bf 74}, 034002 (2007).
\bibitem{QBLi}Q. B. Li, P. N. Shen, Z. Y. Zhang and Y. W. Yu, Nucl. Phys. A {\bf 683}, 487 (2001).
\bibitem{RGM} M. Kamimura, Supp. Prog. Theo. Phys. {\bf 62}, 236 (1977).
\bibitem{Babenko} V. A. Babenko, and N. M. Petrov, arXiv: nucl-th/0307001v1.


\end{thebibliography}
\end{document}